Comment on

# "Unified framework for open quantum dynamics with memory"

by Felix Ivander, Lachlan P. Lindoy and Joonho Lee, Nature Communications 15, 8087 (2024)

by Nancy Makri,[1,2,3,*] Sohang Kundu,[4] Zhenning Cai[5] and Geshuo Wang[5]

[1]Department of Chemistry, University of Illinois, Urbana, Illinois 61801

[2]Department of Physics, University of Illinois, Urbana, Illinois 61801

[3]Illinois Quantum Information Science and Technology Center, University of Illinois, Urbana, Illinois 61801

[4]Department of Chemistry, Columbia University, New York, New York 10027

[5]Department of Mathematics, National University of Singapore, Singapore 119076

[*]email: nmakri@illinois.edu

A recent paper[1] claims to discover the relationship between the generalized quantum master equation (GQME) and the path integral for a system coupled to a harmonic bath. However, this relationship was already established in 2020 by Makri in the context of the small matrix decomposition of the path integral[2,3] (SMatPI). The procedure that Ref. [1] uses in its Supplementary Information (SI) to obtain the various matrices follows steps identical to those of the SMatPI decomposition for the alternative Trotter ordering. The absence of endpoint effects in the kernel matrices of the discretized GQME expression for the reduced density matrix (RDM) is postulated, i.e. it does not follow from and is not consistent with the SMatPI decomposition presented in the SI of Ref. [1] of an auxiliary matrix that differs from the RDM, and gives rise to a crude discrete GQME hierarchy identical to the transfer tensor method[4] (TTM) of Cerrillo and Cao. Further, the Dyck path section of Ref. [1] follows precisely the diagrammatic analysis developed by Wang and Cai in a recent paper. [5] We elaborate on these three critiques in the remainder of this Comment. We use superscripts to indicate corresponding authors' initials (M: Makri, L: Ivander, Lindao and Lee, C: Cerrillo and Cao, W: Wang and Cai). For some expressions we also provide the equation numbers in the relevant articles.

In two papers published in 2020,[2,3] Makri showed for the first time that the path integral expression for the RDM can be exactly decomposed into a sum of small matrix products, eliminating the storage requirements and cost of the tensor-based iterative quasi-adiabatic propagator path integral (QuAPI) algorithm.[6,7] The QuAPI expression[8] for the RDM was based on the following Trotter factorization of the short-time propagator,

$$e^{-i\hat{H}\Delta t/\hbar} \simeq e^{-i\hat{H}_{\text{env}}\Delta t/2\hbar} e^{-i\hat{H}_\text{S}\Delta t/\hbar} e^{-i\hat{H}_{\text{env}}\Delta t/2\hbar} \qquad (1)$$

(hereafter first Trotter splitting) and involves whole- and half-step influence functional (IF) factors but only whole-step system propagators. The SMatPI decomposition[2,3] expressed the $n^2 \times n^2$ RDM matrix $\mathbf{U}_N$ (for all possible initial conditions and with $\mathbf{U}_0 = \mathbf{1}$) at the time $N\Delta t$ in the form

$$\mathbf{U}_N^M = \sum_{r=1}^{N-1} \mathbf{T}_r^M \cdot \tilde{\mathbf{U}}_{N-r}^M + \mathbf{T}_N^M \tag{2}$$

where the auxiliary matrices $\tilde{\mathbf{U}}_k^M$ satisfy

$$\tilde{\mathbf{U}}_k^M = \sum_{r=1}^{k-1} \mathbf{M}_r^M \cdot \tilde{\mathbf{U}}_{k-r}^M + \mathbf{M}_k^M. \tag{3}$$

Here $\mathbf{M}_k^M, \mathbf{T}_k^M, \tilde{\mathbf{U}}_k^M$ are $n^2 \times n^2$ matrices, where $n$ is the number of system states. The subscripts of $\mathbf{M}_r^M$ and $\mathbf{T}_r^M$ indicate time separation in units of the path integral step. (Note that both of these equations can be combined into one[2,3] by labeling the matrices with two subscripts that correspond to the two time points, rather than a single index indicating time difference.) Eq. (3) propagates the auxiliary RDM $\tilde{\mathbf{U}}_k^M$ with the SMatPI midpoint propagation matrices $\mathbf{M}_r^M$ (which are translationally invariant). Using the termination matrices $\mathbf{T}_k^M$ (which differ from $\mathbf{M}_k^M$ through endpoint IF coefficients[9]), Eq. (2) generates the RDM. The need for two sets of matrices is motivated by the path integral expression, where the IF coefficients $\eta_{kk'}$ have different values depending on whether $k = N$ (the time at which the RDM is evaluated) or $k < N$, in close analogy with the use of different edge weights in the trapezoid-rule approximation of an integral. The SMatPI matrices offer the optimal decomposition of the RDM, which minimizes the residual (the last term) at each step. For example, if the IF memory spans a single time step, the path integral variables are connected only to their nearest neighbors, allowing an exact matrix product decomposition without residuals. The SMatPI hierarchy satisfies this requirement, giving $\mathbf{M}_k^M = \mathbf{T}_k^M = 0$ for $k \geq 2$ in this case.

Makri obtained algebraic expressions for the SMatPI matrices $\mathbf{M}_k^M$ and $\mathbf{T}_k^M$ in terms of the system propagators and IF coefficients and proved, analytically as well as numerically, that their elements decrease with increasing index, becoming vanishingly small when the index exceeds the memory length.[2,3] The SMatPI matrices are obtained from equations (2)-(3) within the memory length and are subsequently used for iterative propagation to longer times.

The expression for $\tilde{\mathbf{U}}_k^M$ has a GQME structure (a discrete convolution) with translationally invariant propagation matrices, but different matrices are needed to terminate the hierarchy and obtain $\mathbf{U}_N^M$. This also implies that $\mathbf{M}_1^M \neq \mathbf{T}_1^M \neq \mathbf{U}_1^M$. The loss (by design) of translational invariance at endpoints in the SMatPI treatment (with either Trotter splitting), which offers the optimal decomposition of the path integral expression, is of no consequence to the efficiency of the algorithm, as midpoint and termination matrices are computed simultaneously.[10]

If the propagation and endpoint matrices are assumed equal, one obtains the TTM algorithm[4] of Cerrillo and Cao,

$$\mathbf{U}_N = \sum_{r=1}^{N-1} \mathbf{T}_r^C \cdot \mathbf{U}_{N-r} + \mathbf{T}_N^C, \tag{C-2) or (4}$$

which was introduced as an intuitive *ansatz*, i.e. a dynamical map consistent with the GQME in the $\Delta t \to 0$ limit. The TTM matrices are calculated from Eq. (C-2) using RDM values within memory, which may be computed independently by any method. Eq. (C-2) immediately gives $\mathbf{T}_1^C = \mathbf{U}_1$ and $\mathbf{T}_2^C = \mathbf{U}_2 - \mathbf{U}_1 \cdot \mathbf{U}_1$. The TTM iteration converges to correct results, but (since the TTM algorithm was not derived from the path integral) the hierarchy cannot match the separate IF forms at endpoints and intermediate steps, and thus the matrices are not optimal. For example, the TTM matrices $\mathbf{T}_k^C$ are not zero for $k \geq 2$ even if the IF memory spans a single time step. Referring to the SMatPI decomposition, Makri wrote[3] (omitting citations for clarity): *"This structure is reminiscent of the Nakajima–Zwanzig generalized quantum master equation (GQME), where the time derivative of the RDM depends on the RDM history through a simple time integral. In fact, eq 4.6 bears a close resemblance to the transfer tensor scheme (TTM), which in the $\Delta t \to 0$ limit is equivalent to the GQME. However, the hierarchy obtained through the SMatPI decomposition differs in important ways from the TTM/GQME hierarchy. The latter employs translationally invariant matrices $\mathbf{T}^{(N,N-r)} = \mathbf{T}^{(r0)}$. Such a form appears to be compatible with the path integral only through a crude, unsymmetrized Trotter splitting of the propagators, which is accurate in the $\Delta t \to 0$ limit. The influence functional structure shown in Figure 1 and the SMatPI derivation given in section III clearly show that the small matrix decomposition of the RP must employ different matrices at the endpoints, i.e., $\mathbf{M}^{(r+1,1)} \neq \mathbf{M}^{(r0)}$. The TTM/GQME matrices, which lack this flexibility, have a different structure and larger elements than the SMatPI matrices, i.e., they include a spurious memory and decay slower."* Through this discussion, Makri explicitly established the relationship between GQME and the path integral, while also highlighting important nuances, none of which is acknowledged in Ref. [1]. As stated in Makri's text quoted above, if the IF is replaced by the expression that corresponds to an asymmetric Trotter splitting, the IF endpoint effects are absent, i.e. $\mathbf{I}_{k+m,k} = \mathbf{I}_m, k \geq 1$. Since $\mathbf{I}_{10} = \mathbf{I}_{00} = \mathbf{1}$ in this case, the SMatPI matrices $\mathbf{T}_m^M$ contain IF factors only up to $\mathbf{I}_{m-1}$ and the SMatPI decomposition, Eq. (4.6) of Ref. [3], reverts to the TTM (i.e. discrete GQME) hierarchy with translationally invariant SMatPI matrices which, according to Equations (3.3) and (3.9) of Ref. [3], are given now by

$$T_{1,ij} = U_{1,ij} = I_{0,i} G_{1,ij}$$
$$T_{2,ij} = \sum_k I_{0,i} I_{0,k} \left( I_{1,ik} - 1 \right) G_{1,ik} G_{1,kj} \tag{5}$$

(rewritten here in the notation of Ref. [1] where $\mathbf{G}_1$ is the whole-step forward-backward propagator), etc. In combination with the simple relations between TTM matrices and discretized kernel elements $\Delta t^2 \mathbf{T}_{k+1} = \mathbf{K}_k$ available in [4], these expressions for the SMatPI matrices readily give the GQME kernel elements in terms of IF coefficients. The spurious memory associated with the translationally invariant hierarchy is evident from Eq. (5), which shows that $\mathbf{T}_2 \neq 0$ even if the IF memory spans a single time step.

Ivander, Lindoy and Lee employ[1] a discretized path integral expression for the RDM using the reverse operator ordering,

$$e^{-i\hat{H}\Delta t/\hbar} \simeq e^{-i\hat{H}_S \Delta t/2\hbar} e^{-i\hat{H}_{env} \Delta t/\hbar} e^{-i\hat{H}_S \Delta t/2\hbar} \tag{L-1) or (6}$$

(hereafter second Trotter splitting). As is well known, the choice of operator order in the Trotter factorization is only a matter of convenience and aesthetics. Both orderings give rise to errors of order $\Delta t^3$ and have previously been used in path integral methods. The path integral expression obtained with Eq. (6) contains whole- and half-step system propagators but only whole-step IF factors. In the notation of Ref. [1], it leads to the following expression for the RDM:

$$U^{\text{L}}_{N,x^{\pm}_{2N}x^{\pm}_0} = \sum_{x^{\pm}_{2N-1}}\sum_{x^{\pm}_1} \langle x^+_{2N}|e^{-i\hat{H}_S \Delta t/2}|x^+_{2N-1}\rangle \langle x^+_1|e^{-i\hat{H}_S \Delta t/2}|x^+_0\rangle \langle x^-_0|e^{-i\hat{H}_S \Delta t/2}|x^+_1\rangle$$
$$\times \langle x^-_{2N-1}|e^{-i\hat{H}_S \Delta t/2}|x^-_{2N}\rangle \tilde{U}^{\text{L}}_{N-1,x^{\pm}_{2N-1}x^{\pm}_1} = \sum_{x^{\pm}_{2N-1}}\sum_{x^{\pm}_1} G_{x^{\pm}_{2N}x^{\pm}_{2N-1}}\tilde{U}^{\text{L}}_{N-1,x^{\pm}_{2N-1}x^{\pm}_1}G_{x^{\pm}_1 x^{\pm}_0} \qquad \text{(L-SI-49) or (7)}$$

or, in matrix form,

$$\mathbf{U}^{\text{L}}_N = \mathbf{G} \cdot \tilde{\mathbf{U}}^{\text{L}}_{N-1} \cdot \mathbf{G}, \qquad (8)$$

where $\mathbf{G}$ are forward-backward half-step propagators and $\tilde{\mathbf{U}}^{\text{L}}_{N-1}$ is an auxiliary matrix that involves only whole-step propagators and IF factors with translationally invariant coefficients. Note that endpoint effects are still present with the second Trotter splitting and are included in $\mathbf{U}^{\text{L}}_N$ through the transformation of Eq. (8). The SI of Ref. [1] expresses $\tilde{\mathbf{U}}^{\text{L}}_k$ in the form

$$\tilde{\mathbf{U}}^{\text{L}}_k = \sum_{r=1}^{k-1} \mathbf{M}^{\text{L}}_r \cdot \tilde{\mathbf{U}}^{\text{L}}_{k-r} + \mathbf{M}^{\text{L}}_k \qquad \text{(L-SI-52) or (9)}$$

which (with the choice $\tilde{\mathbf{U}}^{\text{L}}_0 = \mathbf{I}_0$, where $\mathbf{I}_0$ is the matrix of single-time IF factors) involves a single set of propagation matrices. This derivation follows the SMatPI steps, thus Eq. (L-SI-52) is the SMatPI decomposition of the auxiliary matrix $\tilde{\mathbf{U}}^{\text{L}}_k$ with the second Trotter splitting.

Ref. [1] argues that the main advantage of this decomposition is its translationally invariant form, which allows expression of the RDM in GQME form with translationally invariant kernels (i.e. propagation matrices that do not change at endpoints). However, Eq. L-SI-52 applies to the *auxiliary* matrix $\tilde{\mathbf{U}}_k$, which is *not* the actual RDM. To arrive at a GQME form for $\mathbf{U}_N$, Ref. [1] uses a crude GQME discretization to *postulate* a translationally invariant form of the kernel matrices,

$$\mathbf{U}^{\text{L}}_N = \sum_{r=1}^{N-1} \mathbf{T}^{\text{L}}_r \cdot \mathbf{U}^{\text{L}}_{N-r} + \mathbf{T}^{\text{L}}_N, \qquad \text{(L-6) or (10)}$$

where $\Delta t^2 \mathbf{T}^{\text{L}}_{k+1} = \mathbf{K}^{\text{L}}_k$ (the discretized GQME kernel element) for $k > 1$. Eq. (10) is identical to Cerrillo and Cao's Eq. (2), i.e., $\mathbf{T}^{\text{L}}_k = \mathbf{T}^{\text{C}}_k$. Thus, the propagation hierarchy proposed by Ivander, Lindoy and Lee is precisely the TTM algorithm. However, this important connection is not stated in Ref. [1] either.

Eq. (L-6) for the RDM $\mathbf{U}^{\text{L}}_N$ does not follow from the decomposition of the auxiliary RDM $\tilde{\mathbf{U}}^{\text{L}}_k$, i.e., the TTM matrices are not given by applying the transformation of Eq. (8) to the $\mathbf{M}^{\text{L}}_r$ matrices obtained in Eq. (L-SI-52) (note the extra terms). The translational invariance of the GQME kernel elements obtained in Ref. [1] is an *ansatz* suggested by the coarse discretization of the integrodifferential equation and does not arise from the translational invariance of the $\tilde{\mathbf{U}}^{\text{L}}_k$ decomposition attained through the second Trotter factorization in the path integral expression. The implication in Ref. [1] of a translationally invariant decomposition of the path integral achieved through the second Trotter splitting is not substantiated by the expressions presented in the article and the SI.

By applying to the postulated Eq. (L-6) steps analogous to those used in the SMatPI derivation,[2,3] Ivander, Lindoy and Lee obtain in their SI expressions for the TTM matrices and kernel elements, equations (L-7)-(L-10). Since the SMatPI and TTM hierarchies (with either Trotter factorization) are similar in structure, the TTM matrices and GQME kernel elements obtained in Ref. [1] bear a close similarity to Makri's SMatPI matrices, although the TTM matrices contain additional terms that amount to spurious memory resulting from the use of fixed matrices, as discussed in the paragraph quoted from Ref. [3]. This similarity is made evident by comparing the diagrams presented in Fig. 1 of Ref. [1] to those in Fig. 2 of Ref. [2]. Ref. [1] also presents in its Fig. 2 the decay in magnitude of the matrices as a function of time. The decay is very similar to that presented in Fig. 3 of Ref. [2] and Fig. 2 of Ref. [3].

We also note that by applying the SMatPI decomposition to the QuAPI expression for Hamiltonians augmented with time-dependent fields,[11] Makri obtained a small matrix path integral algorithm for driven system-bath dynamics that avoids the use of tensors.[12] The relevant section of Ref. [1] does not mention the existence of such a method. Ref. [1] also proposes storing propagation tensors to avoid re-computing the propagation matrices. However, proceeding this way, the method becomes similar to the iterative QuAPI algorithm, which has been used successfully in many investigations of driven dissipative systems, its main limitation being the required tensor storage.

Last, Wang and Cai[5] recently analyzed Makri's SMatPI expression[2,3] and showed that the terms in the SMatPI matrices are represented by Dyck paths and that the cardinality of each set is given by the Catalan number. The correspondence between terms in a SMatPI matrix and Dyck paths is pointed out in Section 4 of Ref. [5], and more intuitively by Figures 5 and 6 (which are the same as Figures 4 and 5 in the preprint version cited by Ref. [1]). Since the number of Dyck paths is given by the Catalan number, the number of terms in the SMatPI matrices can be determined using basic combinatorics. The kernel elements in Ref. [1] share the same structure. For example, Eq. (L-9) and Eq. (L-10) have identical structure as Eq. (W-3.6) and Eq. (W-4.2), respectively. It is thus unsurprising that the kernel elements in Ref. [1] are described by Dyck paths. This common structure further demonstrates the similarity of the two (SMatPI and TTM/crudely discretized GQME) hierarchies and algorithms, which differ in the details of system propagator factors and IF coefficients. The section on Dyck paths in Ref. [1] does not credit Wang and Cai for this development, and has no mention of their existing work establishing similar expressions based on the framework of SMatPI, merely citing a preprint version in reference to the Catalan number even though the preprint version of Ref. [5] includes sufficient details about the Dyck paths.

In summary, the main developments presented in Ref. [1] bear very close resemblance to published work. To provide further perspective, we discuss briefly the desirable structure of discretized GQMEs. While the simplest form is a sum involving translationally invariant kernel matrices, that treatment is consistent with the crudest approximation to the time derivative and integral, with error that scales as $\Delta t^2$. It is not surprising that such a structure is compatible with a path integral expression in terms of unsymmetrized Trotter propagators (which also have error of order $\Delta t^2$). One also expects a symmetrized Trotter path integral expression, which is characterized by $\Delta t^3$ error (thus offering higher accuracy) to be consistent with an improved discretization of the integrodifferential GQME that would involve different matrices at the endpoints. Constructing the matrices from the path integral offers this flexibility. The optimal SMatPI decomposition of the RDM with symmetrized short-time propagators has precisely this structure and can be thought of as an improved discretization of the GQME.

**Acknowledgments**

This material is based upon work supported by the National Science Foundation under Award CHE-1955302.

**Author Contributions**

NM wrote the paper. SK, ZC and GW edited the paper.

**Competing Interests**

The authors declare no competing interests.